\documentclass[article, aps,pra,twocolumn,unsortedaddress,superscriptaddress]{revtex4-2}
\usepackage{hyperref}
\usepackage{graphicx}
\usepackage{dcolumn}
\usepackage{bm}
\usepackage{mathtools}
\usepackage{amsmath}
\usepackage{amssymb}
\usepackage{multirow}
\usepackage{xcolor}

\newcommand{\ii}{\mathrm{i}}

\newcommand{\uss}{\uparrow}
\newcommand{\dss}{\downarrow}

\begin{document}

\title{Quantum walk on a square lattice with identical particles}
\author{Gonzalo Camacho}
\email{gonzalo.camacho@dlr.de}
\affiliation{%
  Department High-Performance Computing, Institute of Software Technology, German Aerospace Center (DLR), 51147 Cologne, Germany}
 
\author{Jasmin Meinecke}
\affiliation{Institute of Physics and Astronomy, Technische Universit\"{a}t Berlin, 10623 Berlin, Germany}
\affiliation{Max-Planck-Institut f\"{u}r Quantenoptik, Hans-Kopfermann-Stra{\ss}e 1, 85748 Garching, Germany}
 
 \author{Janik Wolters}
  \affiliation{Institute of Physics and Astronomy, Technische Universit\"{a}t Berlin, 10623 Berlin, Germany}
 
\affiliation{%
 Institute of Optical Sensor Systems, German Aerospace Center (DLR), 12489 Berlin, Germany}

 \affiliation{Einstein Center Digital Future (ECDF), Wilhelmstraße 67, 10117 Berlin, Germany}
 
 \affiliation{AQLS UG (haftungsbeschränkt), Guerickestraße 12, 10587 Berlin, Germany}

\date{\today}
\begin{abstract}
    We investigate quantum superposition effects in two-dimensional quantum walks of identical particles with different statistics under particle exchange, starting from various different initial configurations. To characterize interparticle correlation dynamics, we focus on joint properties such as two-particle coincidence probabilities and the spread velocity of the interparticle distance. Regarding spatial modes as an environment for the particles internal degrees of freedom, we study the role played by the particle statistics using standard entanglement witnesses, showing that particles possessing fermionic statistics are more resistant to thermalize with their environment. We analyze the presence of multipartite entanglement in the system's degrees of freedom through the Quantum Fisher Information, revealing that fermionic states generated during the walk are better suited to perform quantum metrology tasks. Finally, we discuss the potential for implementing this model using integrated photonic circuits by exploiting $N$-partite entanglement between individual photons. 
\end{abstract}

\maketitle

\section{Introduction}

Quantum walks~\cite{Aharonov1993,Kempe2003a,qiang2024review}(QWs) have gained considerable attention as a candidate model for achieving universal quantum computation~\cite{Childs2009,Lovett2010,Childs2013}. Among many other relevant applications, QWs have revealed the potential for achieving quantum advantage respect to certain classical search-based~\cite{Shenvi2003,Childs2004,Santha2008,Tulsi2008} and sampling~\cite{Qiang2016} algorithms when implemented on quantum processors.

Due to their versatility, QWs have been explored in a wide range of scenarios, notably focusing on two-particle correlations of QWs on a line~\cite{Knight2003,Stefanak2006,Pathak2007,Stefanak2011,Liu2009,Rigovacca2016}, stochastic QWs~\cite{Whitfield2010,Caruso2014,Govia2017,Schuhmacher2021} and multiparticle QWs~\cite{Rohde2011,Costa2019,Li2019,Ostahie2023}, to name a few. The dimensionality of space where walkers are allowed to propagate plays a pivotal role in the properties of the QW~\cite{Mackay2002}. While QWs restricted to one dimensional (1D) geometries have been widely studied, two dimensional QWs (2DQWs) offer a much richer landscape of geometries to explore interparticle interference effects, and are known to be in close connection with Grover's search algorithm~\cite{Grover1997,Watanabe2008,DiFranco2011,DiFranco2011a}. 

In recent decades, QWs on integrated photonic circuits~\cite{AspuruGuzik2012,Graefe2016,Graefe2020,Wang2020} were one of the first testbeds for photonic quantum information processing~\cite{OBrien2009,Flamini2018,Slussarenko2019,Pelucchi2022}. This approach has been supported by the great experimental success witnessed in preparing entangled multiphoton states~\cite{Bouwmeester1999,Mitchell2004,Barreiro2005,Matthews2009,Pan2012,Yao2012,Wang2016,Malik2016,Reimer2016,Thomas2022}, where experiments involving photonic QWs have mostly focused on individual or polarization-entangled photon states~\cite{Peruzzo2010,Schreiber2012,Kitagawa2012,Sansoni2012,Poulios2014,Pitsios2017,Tang2018,Wang2019,Jiao2021,Esposito2022,Lin2023}. Due to their relatively long coherent times and the high degree of experimental control in the study of photon interference effects~\cite{Hong1987}, photonic processors are now being considered as good candidates for developing quantum computers~\cite{Noh2016} capable to simulate complex quantum systems~\cite{Chen2021,Sparrow2018,Somhorst2023}. A promising application route for QW computing in integrated photonics is to develop photonic quantum simulators that mimic quantum interference effects in systems of identical particles~\cite{Matthews2013,Carolan2014,Menssen2017}, allowing one to observe the role played by particle statistics during the evolution of the walk~\cite{Omar2006,Brennen2010,Lehman2011,Chandrashekar2012QuantumWO,Wang2014b,Cai2021,Lau2022}. Additionally, the potential to realize entangled multiphoton states~\cite{Greenberger1990,Zeilinger1998} employing photonic quantum processors opens up the possibility to explore fundamental questions about quantum entanglement~\cite{Brunner2014,Horodecki2009} in systems of identical particles~\cite{Reusch2015,Wasak2018,Benatti2020}.   

A setup where correlations between identical particles can be extracted using photonic quantum simulators are discrete QWs, where at each step of the walk a coin operation entangles the particles internal degrees of freedom and their spatial modes. In Ref.~\cite{Omar2006}, a 1D, discrete-time QW of two entangled particles having either boson or fermion statistics was investigated, concluding that particle statistics have a significant effect on joint properties during the evolution of the walk. A natural extension of these results, which constitutes the main motivation of this work, is to consider the situation corresponding to a less spatially constrained geometry, while increasing the number of walkers and allowing for the particles to have intermediate exchange statistics ranging between bosons and fermions. 

For particles prepared in a spatially constrained state, subsequent steps of the walk will spread the wavefunction through different spatial modes, which are regarded as an environment. Another interesting question to address in this context is how particle statistics affect the dynamics of entanglement between the particles internal degrees of freedom (system) and their spatial coordinates (environment), with the goal to identify genuine entangled states resisting environment thermalization. In this context, thermalization should be understood as how entangled the system degrees of freedom become with the environment when the walkers propagate over an infinitely extended spatial geometry.

Generating states whose system dynamics are nearly isolated from the environment is of interest to perform quantum metrology tasks. In particular, recent works on 1D QWs of identical particles have reported the essential role played by particle statistics in the Quantum Fisher Information (QFI)~\cite{Cai2021}, which characterizes the maximum achievable precision in phase parameter estimation under unitary evolution. The direct relevance of the QFI to identify resourceful states that can be generated by QWs is evident, given that the QFI also is a genuine multipartite entanglement witness having well-defined entanglement bounds~\cite{Toth2012}.

The above questions are motivated not only from a fundamental point of view, but also regarding the experimental feasibility of employing photonic quantum computing platforms as simulators for systems of identical particles, and how do these compare against an implementation on alternative quantum processor architectures, e.g. on superconducting qubits hardware. The proposal to employ multilevel, multipartite entanglement in integrated photonic circuits of QWs~\cite{Matthews2013} sets the basis to address these questions employing linear optics, with time-bin encoding~\cite{Jayakumar2014,Lubasch2018} methods offering a promising route to realize the required multipartite scenario of the photons. 

In this work we address the QW of a state representing \emph{identical} quantum particles allowed to move on a 2D square lattice geometry. We study this construction using a first quantization scheme, which naturally employs a multipartite setting of distinguishable particles for constructing the full Hilbert space. The influence of the particle statistics is analyzed in two different settings. First, with the goal to characterize interparticle correlation dynamics, we develop an efficient method for calculating joint properties from a special set of initial states, allowing us to perform detailed numerical simulations employing different initial configurations of the walkers. In particular, we focus on the dynamics of two-particle coincidence events, which is a quantity of experimental relevance~\cite{Matthews2013,Carolan2014,Menssen2017} and the two-particle spread velocity~\cite{Omar2006}, which is of interest from a theoretical point of view in QWs with identical particles. Second, in the context of system-environment thermalization, we conduct further numerical simulations by looking at standard entanglement witnesses between the spatial and the internal degrees of freedom of the walkers. In particular, we choose the von Neumann entropy to characterize the entanglement between spatial and internal degrees of freedom, and the purity of the reduced density matrix of the system as a measure of mixedness directly related to separability of the state. In order to identify favourable statistics of the particles to perform quantum metrology tasks and to characterize multipartite entanglement dynamics during the walk, we look at the QFI. Since we are dealing with identical particles and the QFI depends on the particular choice of the operator, a particularly relevant scenario is to consider quantum metrology in the class of Ising-like Hamiltonians.

The paper is structured as follows. In Sec.~II, we introduce the QW model, describing the Hilbert space structure, the initial state and the unitary evolution operator. We also comment on the feasibility to experimentally simulate states describing identical particles employing linear optics elements in this section. In Sec.~III, we give a brief description of the main observables studied in this work, which includes joint two-particle properties, entanglement witnesses and a generalized Bell inequality. In Sec.~IV we present the main results. Sec.~V concludes with a summary and we discuss open questions.

\begin{figure}[!t]
    \centering
    \includegraphics[scale=0.6]{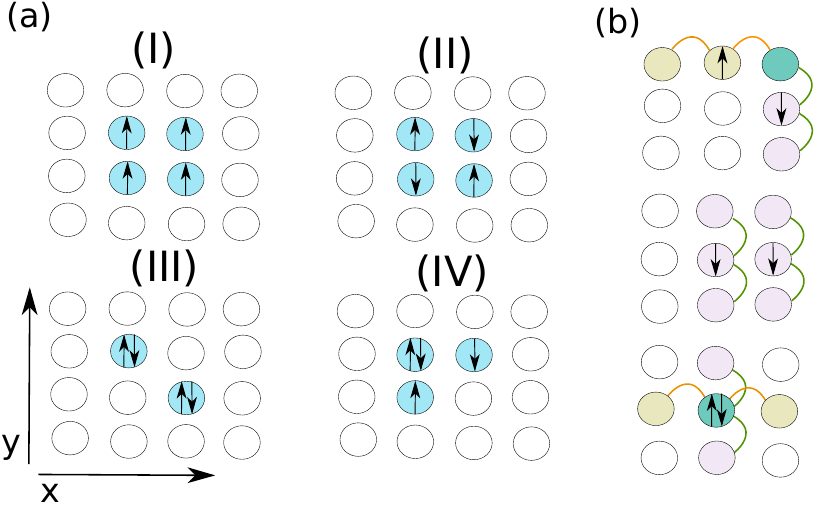}
    \caption{(a) Different initial state configurations for the state $|\Psi_0\rangle$ defined in Eq.~\eqref{eq:psi0}, for $N=4$ particles on a square lattice of $L\times L$ lattice sites with $L=4$. Occupied spatial modes allocating particles with $\sigma=\uparrow,\downarrow$ internal degrees of freedom are filled in blue, whereas empty spatial modes are represented by a white background circle. The initial state $|\Psi_0\rangle$ incorporates different particle statistics through the phase $\phi$. (b) Illustration of the conditional hopping process described in Eq.~\eqref{eq:uni_def} for initial product states in the $|x,y,\sigma\rangle$ basis; hopping between nearest neighbors have been represented with lines, and allowed spatial modes due to hopping processes have been colored. The green regions represent spatially shared modes due to shared hopping into that lattice position.}  
    \label{fig:fig1}  
\end{figure}
\section{Model}
\subsection{Hilbert space}
We consider $N$ distinguishable quantum particles, each labeled by an index $j=1,...,N$. The quantum state associated with particle $j$ is $|\psi_j\rangle$, living in a finite dimensional Hilbert space $\mathcal{H}_j$; the total Hilbert space of the whole system is $\mathcal{H}=\mathcal{H}_1\otimes \mathcal{H}_2\otimes...\otimes \mathcal{H}_N$. We consider the Hilbert space of the $j$-th particle to be the tensor product of the Hilbert space $\mathcal{S}_j$ associated with spatial coordinates of the particle, and the space $\mathcal{C}_j$ associated with their internal degree of freedom:
\begin{eqnarray}
\mathcal{H}_j=\mathcal{S}_j\otimes \mathcal{C}_j.
\end{eqnarray}
We consider all $\mathcal{H}_j$ to be equivalent regardless of the $j$ index. We restrict the spatial geometry in $\mathcal{S}_j$ to a finite 2D square lattice of dimension $L\times L$ with discrete spatial coordinates $\Lambda:(x,y)|x,y\in\{0,1,...,L-1\}$; arbitrary positions in the lattice $\Lambda$ are represented by a vector $\vec{r}=(x,y)$. The subspace $\mathcal{C}_j$ has local basis states $|\sigma\rangle\in \{|\uparrow\rangle,|\downarrow\rangle \}$, which will be regarded as a spin-$1/2$ degree of freedom. The basis of states $\{|x,y,\sigma\rangle\}$ spans any $\mathcal{H}_j$, with total dimension $d=\text{dim}(\mathcal{H}_j)=2L^2$; the total Hilbert space dimension $\text{dim}(\mathcal{H})=d^N$. In what follows, we will often refer to the Hilbert space $\mathcal{C}=\mathcal{C}_j^{\otimes N}$ associated only with the particles internal degrees of freedom as the \emph{system}, while interpreting the Hilbert space of spatial coordinates $\mathcal{S}=\mathcal{S}_j^{\otimes N}$ as the \emph{environment}.

\subsection{Initial state}\label{subsec:initial_state}

We consider states describing $N$ \emph{identical} particles of the form:
\begin{eqnarray}\label{eq:psi0}
|\Psi_0\rangle=\frac{1}{\sqrt{N!}}\sum_{\mathcal{P}}e^{\ii\phi_{\mathcal{P}}}| \psi^0_{1_\mathcal{P}}, ...,\psi^0_{j_\mathcal{P}},..., \psi^0_{N_\mathcal{P}} \rangle,
\end{eqnarray}
where $\mathcal{P}$ labels a single element of the set containing all permutations of $I_N=\{1,...,j,...,N\}$, with $\mathcal{P}=\{ 1_\mathcal{P},...,j_\mathcal{P},...,N_\mathcal{P}\}\in S_N(I_N)$, and $j_\mathcal{P}\in I_N,\forall j$. The phase $\phi_{\mathcal{P}}=0$ if the number of permutations associated with $\mathcal{P}$ is even, whereas $\phi_{\mathcal{P}}=\phi\in[0,\pi]$ if the number of permutations is odd. For $\phi=0$, Eq.~\eqref{eq:psi0} represents a state of particles obeying boson statistics, whereas for $\phi=\pi$ the state in Eq.~\eqref{eq:psi0} corresponds to particles obeying fermion statistics. The single parameter $\phi$ tunes continuously between the two. For the special cases $\phi=0,\pi$, the state is invariant under arbitrary particle pair exchange. For $\phi\in(0,\pi)$, the state picks up a nonglobal phase under particle exchange. 

For the initial conditions we consider product states $|\psi_j^0\rangle^{\otimes j}$ formed from a set of $N$ states $S_0=\{|\psi_1^0\rangle,...,|\psi_j^0\rangle,...,|\psi_N^0\rangle\}$; we choose each $|\psi_j^0\rangle$ to be an element of the $\{|x,y,\sigma\rangle\}$ basis, so that all states in $S_0$ are pairwise orthogonal. The initial state is constructed using the $N!$ permutations from the set $S_0$ as in Eq.~\eqref{eq:psi0}. Since the state in Eq.~\eqref{eq:psi0} describes identical particles, it is not possible to assign to any of the particles a definite initial condition from $S_0$. 

We will consider four different initial configurations of the particles, depicted in Fig~\ref{fig:fig1}. Configurations (I) and (II) correspond to the case where the walkers initially do not share any spatial modes; however, the two situations are rather different in that (I) has the spatial degrees of freedom completely decoupled from the internal ones (any particle subspace has a $\uparrow$ internal degree of freedom), whereas (II) is not separable from the environment due to the internal degrees of freedom being different. In other words, spatial and internal degrees of freedom in (II) are expected to be entangled, whereas in (I) this is not the case. In configurations (III) and (IV) we address the situation where at least two spatial modes are initially shared. In (III), if spatial degrees of freedom are traced out, the internal degrees of freedom for each particle pair are highly entangled as in the case of Bell-like states (for each filled spatial mode) including a relative phase $\phi$. In contrast, in (IV) we break this symmetric scenario by leaving a single spatial mode to be shared between the walkers.

\subsection{Unitary evolution operator}\label{subsec:unitary_evol}
The dynamics of the state are determined by a unitary operator $\hat{U}$ acting on the composite Hilbert space of particles. We consider a total unitary given by:
\begin{eqnarray}
\hat{U}=U^{\otimes N},
\end{eqnarray}
i.e. each unitary $U$ acts independently on each particle subspace. We consider a model for a discrete time QW (DTQW) on the square lattice geometry $\Lambda$, with $U$ given by:
\begin{eqnarray}\label{eq:uni_def}
U&=&U_{\mathcal{S}}U_{\mathcal{C}}\nonumber\\
&=&\underbrace{\left[e^{-\ii H_{\parallel}}\otimes|\uparrow\rangle\langle \uparrow| + e^{-\ii H_{\perp}}\otimes|\downarrow\rangle\langle \downarrow|\right]}_{=U_{\mathcal{S}}}\nonumber\\
&\times & \underbrace{I_{\mathcal{S}}\otimes\left[\cos(\theta)(P_{\uparrow\uparrow}-P_{\downarrow\downarrow})+\sin(\theta)(P_{\uparrow\downarrow}+P_{\downarrow\uparrow})\right]}_{=U_{\mathcal{C}}},
\end{eqnarray}
where we have defined $P_{\sigma,\sigma'}=|\sigma\rangle\langle\sigma'|$, with $\sigma,\sigma'\in\{\uparrow,\downarrow\}$. The unitary $U_{\mathcal{C}}$ acts on the internal degrees of freedom of particles (with $I_{\mathcal{S}}$ being the identity for the $\mathcal{S}$ subspace) as a coin operation depending on the parameter $\theta$; in what follows, we fix $\theta=\pi/4$. The unitary $U_\mathcal{S}$ is a conditional step operator, where $H_{\perp},H_{\parallel}$ represent two nearest-neighbour hopping Hamiltonians along vertical and horizontal directions of the lattice, respectively. The conditional shift process is then as follows: If the internal degree of freedom of the walker is in the $|\downarrow\rangle$ state, then the walker evolves under a unitary operator with Hamiltonian given by $H_\perp$, which contains only nearest-neighbor hoppings along the vertical direction of the lattice; otherwise, if the internal state is $|\uparrow\rangle$, the Hamiltonian governing the step includes hoppings along the horizontal directions of the lattice. In the basis of $\mathcal{S}$, the exponents are the Hermitian operators given by:
\begin{eqnarray}\label{eq:hams}
H_{\perp}&=& t\sum_{x,y}|x,y+1\rangle \langle x,y|+\text{H.c.},\nonumber\\
H_{\parallel}&=& t\sum_{x,y}|x+ 1,y \rangle\langle x,y|+\text{H.c.}.
\end{eqnarray}
The parameter $t$ tunes the hopping amplitude. A small value of $t$ corresponds to a nearly adiabatic evolution of the walkers after application of $\hat{U}$, with the emergent interference effects due to different particle statistics becoming more evident compared to the case  $t=1$, which will contribute having amplitudes between very distant spatial points. Unless otherwise stated, for all numerical results we consider open boundary conditions in the lattice, e.g., $H_\perp$ contains a term $|x,L\rangle\langle x,L-1|+\text{H.c.}$, but not a term $|x,L\rangle\langle x,0|+\text{H.c.}$. In more detail, we always make sure that for the system sizes and the value of the parameter $t$ employed, boundaries are never reached (see Fig.~\ref{fig:app2} in Appendix~\ref{app:app2}). This way, the situation is equivalent to particles evolving in an infinitely extended spatial grid. When periodic boundary conditions are considered, the model enjoys translational symmetry and can be solved exactly. The exact eigenvalues and eigenvectors of $U$ for the case of periodic boundary conditions are included in Appendix~\ref{app:app1}. We note that in Eq.~\eqref{eq:uni_def}, $\hat{U}$ involves complex amplitudes.

The model presented in Eq.~\eqref{eq:uni_def} differs significantly from standard QW models found in the literature, in particular regarding the so called split-step walks that are relevant for the exploration of topological properties~\cite{Kitagawa2010,Edge2015}. For comparison purposes, we have included results corresponding to a split-step walk in Appendix~\ref{app:app4}. One of the main motivations to choose such a model for states given by Eq.~\eqref{eq:psi0} is that it introduces changes in the relative complex phases during the walk, while keeping the binary nature of the conditional step operator. The model in Eq.~\eqref{eq:uni_def} belongs to a class of models where the standard left or right shift operators present in 1D QWs are substituted by more generic unitaries, whose associated Hamiltonian might have a physical origin. In this sense, the model describes a situation reminiscent of systems experiencing time-reversal symmetry breaking, with opposite internal degrees of freedom evolving under different Hamiltonians. A physically relevant scenario of such situation is that of 2D systems in the presence of magnetic fields, which can also be realized with polarized beam splitters in optics.

The quantum state after $n$ steps of the walk is given by:
\begin{eqnarray}
|\Psi(n)\rangle &=& \hat{U}^n|\Psi_0\rangle\nonumber\\
&=&\frac{1}{\sqrt{N!}}\sum_{\mathcal{P}}e^{\ii\phi_{\mathcal{P}}}U^n|\psi^0_{1_\mathcal{P}}\rangle\otimes...\otimes U^n|\psi^0_{N_\mathcal{P}}\rangle.
\end{eqnarray}

\subsection{Multi-partite entanglement over replica configurations employing linear optics}\label{subsec:multi_gene}

Recent experimental advances in integrated photonic architectures have shown how to exploit multilevel, multipartite entanglement in order to generate states of identical particles with arbitrary exchange statistics employing elements from linear optics~\cite{Matthews2013}, i.e. single photons, providing these are entangled beforehand (which constitutes itself an experimental challenge). The key idea is to employ $N$ replicas of a given mode transformation (any unitary evolution operator for the particles), one for each of the photons, so that each replica subspace (regarded as the $\mathcal{H}_j$ subspaces) contains one particle. Each of the particles of the replicas is entangled to each other employing an experimentally controllable parameter $\phi$ for simulating states with different statistics upon particle exchange; that is, simulating states of \emph{indistinguishable particles}. We stress that although the particles employed in the experiment to construct such state have always well defined statistics (in the case of photons having bosonic nature), the above state can be potentially realized in experiments for instance, employing time-bin encoding of photons. This constitutes the fundamental basis for creating a quantum simulator employing single photons~\cite{Matthews2013}, the requirement being the preparation of entangled-photon states in the multipartite setting scenario.

For QWs consisting of two walkers, polarization degrees of freedom of the photons is sufficient to recreate identical particle statistics states~\cite{Matthews2013}. For more than two walkers, additional degrees of freedom are needed to generate the multipartite setting. In this direction, photonic time-bin encoding~\cite{Jayakumar2014,Lubasch2018} presents a promising route to realize the required encoding over extended Hilbert spaces.

\section{Observables}\label{sec:observables}
 
\subsection{Two particle distribution}\label{subsec:twoparticle}
A quantity commonly explored in multiparticle QW experiments~\cite{Peruzzo2010,Sansoni2012,Matthews2013,Poulios2014} is the two-particle correlation at arbitrary points $\vec{r}_1,\vec{r}_2$ in the lattice $\Lambda$. They are defined at the $n$-th step of the walk by:
\begin{eqnarray}\label{eq:c12_def}
C_{\sigma_1,\sigma_2}(\vec{r}_1,\vec{r}_2,n)&=&\langle\Psi(n)|\hat{P}_{i,\vec{r}_1\sigma_1}\hat{P}_{j,\vec{r}_2\sigma_2}|\Psi(n)\rangle,
\end{eqnarray}
where $i,j$ indicate any pair of particle subspaces, and $\hat{P}_{i,\vec{r}_1\sigma_1}$ is the projector for particle $i$ at lattice site $\vec{r}_1$ and internal degree $\sigma_1$. Since $|\Psi(n)\rangle$ describes identical particles, the choice of $i,j$ is arbitrary. The correlation $C_{\sigma_1,\sigma_2}(\vec{r}_1,\vec{r}_2,n)$ gives the \emph{joint probability density} for any pair of particles. Assuming orthogonal states in the set $S_0$, these distributions can be calculated using the approach described in Appendix~\ref{app:app2}.

The explicit expression for the joint probability density distribution is given by:
\begin{eqnarray}\label{eq:probdens}
P(\vec{r}_1,\vec{r}_2,n)&=&C_{\uparrow,\uparrow}(\vec{r}_1,\vec{r}_2,n)+C_{\downarrow,\downarrow}(\vec{r}_1,\vec{r}_2,n)\nonumber\\
&+&C_{\uparrow,\downarrow}(\vec{r}_1,\vec{r}_2,n)+C_{\downarrow,\uparrow}(\vec{r}_1,\vec{r}_2,n).
\end{eqnarray}
The probability of a two-particle coincidence event at time step $n$ can be obtained from Eq.~\eqref{eq:probdens} as:
\begin{eqnarray}\label{eq:diag_prob}
    P_{\text{diag}}(n)=\sum_{\vec{r}}P(\vec{r},\vec{r},n).
\end{eqnarray}

\subsection{Average inter-particle distance}\label{subsec:av_speed}
The average interparticle distance $\Delta_{12}(n)$ is defined as:
\begin{eqnarray}\label{eq:delta12}
\Delta_{12}(n)&=&\sum_{\vec{r}_1,\vec{r}_2}\sqrt{(x_1-x_2)^2 + (y_1-y_2)^2} P(\vec{r}_1,\vec{r}_2,n),\nonumber\\
\vec{r}_1&=&(x_1,y_1),\hspace{5pt}\vec{r}_2 =(x_2,y_2).
\end{eqnarray}
This quantity represents the average separation of any two particles in the system at any step $n$ of the walk. Since this is a joint property, knowledge of the full joint distribution $P(\vec{r}_1,\vec{r}_2,n)$ is required. This quantity has been reported to scale linearly with the number of steps $n$ for a QW of two entangled particles on a 1D line~\cite{Omar2006}. 

If linear behavior of $\Delta_{12}(n)$ is expected in the limit $L\to\infty$, the asymptotic behavior is characterized by a spreading parameter:
\begin{eqnarray}\label{eq:vf_phi}
v_\phi=\frac{\Delta_{12}(\tau)}{\tau}, \hspace{4pt}\tau=n t,
\end{eqnarray}
where $t$ is the hopping parameter in Eq.~\eqref{eq:hams}, and $v_\phi$ is the average separation speed between particles dependent on the phase $\phi$, i.e., the particle statistics.

\begin{figure}[!t]
    \centering
    \includegraphics[scale=0.68]{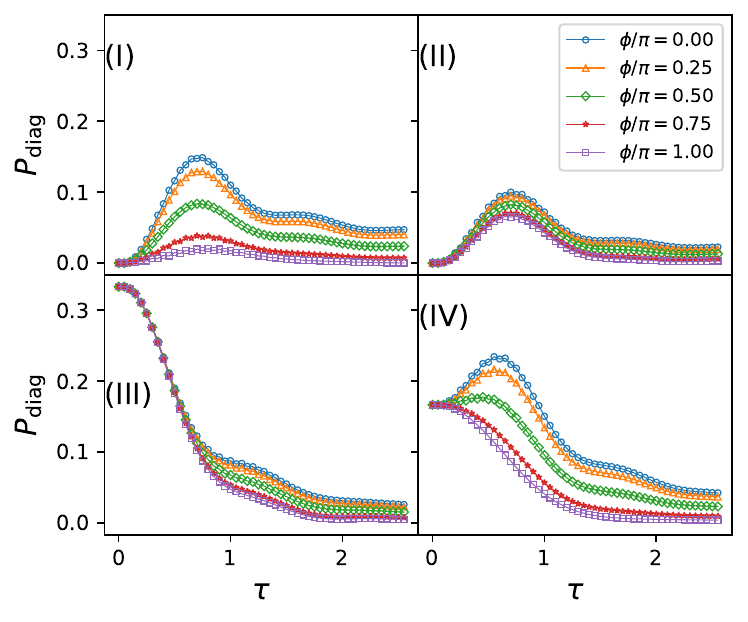}
    \caption{The two-particle coincidence probability as defined in Eq.~\eqref{eq:diag_prob}, for different values of the particle exchange statistics parameter $\phi$. Bosons tend to occupy equal spatial modes during the evolution, leading to pronounced peaks in the probability of coincidence events. For fermions, such coincidences are less likely due to the Pauli exclusion, although peaks might still develop as in (II). Data corresponds to the different initial conditions (I), (II), (III) and (IV) from Fig.~\ref{fig:fig1}, with $L=16$, $t=0.05$ and a total of $n=50$ time steps, with $\tau=nt$. }
    \label{fig:fig2}
\end{figure}

\begin{figure}[!t]
    \centering
    \includegraphics[scale=0.66]{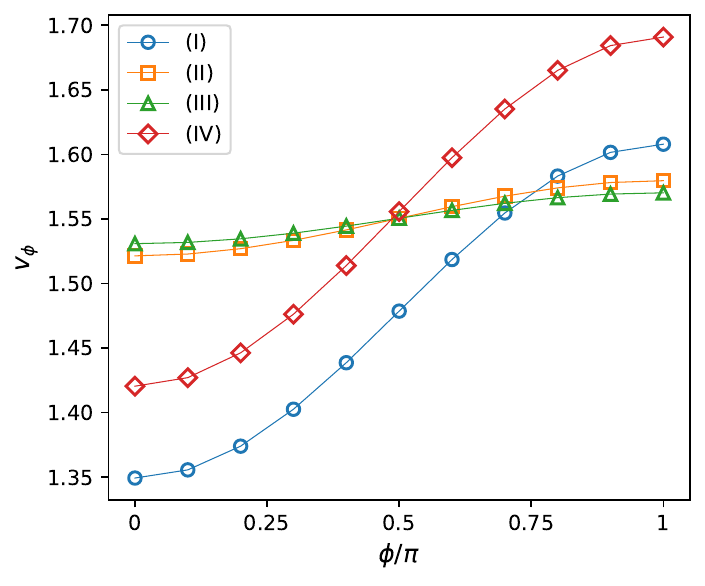}
    \caption{Average speed repulsion $v_\phi$ for the interparticle distance $\Delta_{12}$ as defined in Eq.~\eqref{eq:vf_phi}, for different values of the statistical parameter $\phi$ and number of walkers $N=4$ using the initial conditions (I),(II),(III) and (IV) from Fig.~\ref{fig:fig1}. As expected, particles with fermion statistics ($\phi=\pi$) tend to repel each other faster than bosons ($\phi=0$) as a consequence of the Pauli exclusion principle. The value of $v_\phi$ depends on the initial configuration of particles, sharing qualitative behavior in (I) and (IV) [(II) and (III)]. We used $L=40$ in all cases. }
    \label{fig:fig3}
\end{figure}


\subsection{Quantum Fisher Information, entanglement entropy and purity}\label{subsec:ent_meas}
The Quantum Fisher Information~\cite{Pezze2018} (QFI) of a general state $\rho$ respect to the observable $\mathcal{O}$ is given by:
\begin{eqnarray}\label{eq:qfi_def}
    \mathcal{F}_Q[\rho,\mathcal{O}]=2\sum_{k,l}\frac{(\lambda_k-\lambda_l)^2}{\lambda_k+\lambda_l}|\langle k|\mathcal{O}|l\rangle|^2,
\end{eqnarray}
where $\lambda_k$ and $|k\rangle$ are the eigenvalues and eigenvectors of $\rho$, respectively. Tracing out all spatial degrees of freedom, we consider the reduced density matrix of the particles internal degrees of freedom at any time step $n$:
\begin{eqnarray}\label{eq:rhoc_state}
    \rho_\mathcal{C}=\text{Tr}_{\mathcal{S}}(|\Psi(n)\rangle\langle\Psi(n)|),
\end{eqnarray}
where we omit the explicit step dependence in $\rho_\mathcal{C}$ for notation convenience. Since we are dealing with a state of identical particles in $\rho_\mathcal{C}$ and we are interested in joint properties between any particle pairs, a multipartite operator $\mathcal{O}\in \mathcal{C}^{\otimes N}$ where particle statistics effects are expected to be of relevance is the uniform Ising Hamiltonian:
\begin{eqnarray}\label{eq:hop_qfi}
    \mathcal{O}=\sum_{j< k}\sigma_j^z\sigma_k^z,
\end{eqnarray}
where $\sigma_j^z$ is the $z$-Pauli matrix for the $j$-th particle subspace index. 

As a witness of bipartite entanglement between the system and the environment, we employ the von Neumann entropy for the reduced density matrix $\rho_\mathcal{C}$:
\begin{eqnarray}\label{eq:vn_entropy}
    S_\mathcal{C}=-\text{Tr}\left(\rho_\mathcal{C}\log_2(\rho_\mathcal{C})\right).
\end{eqnarray}

The degree of mixedness of $\rho_{\mathcal{C}}$ with the environment can be obtained from the purity:
\begin{eqnarray}\label{eq:purity}
    P_\mathcal{C}=\text{Tr}(\rho_\mathcal{C}^2).
\end{eqnarray}
The purity is related to the linear entropy, a lower approximation of the von Neumann entropy~\eqref{eq:vn_entropy}. For a system of $N=4$ walkers, $\rho_\mathcal{C}$ has dimension $16\times 16$, and thus $\text{min}\left(P_\mathcal{C}\right)=\frac{1}{16}$.

\subsection{Bell test}\label{subsec:bell_test}

Bell inequalities establish bounds on the classical nature of correlations in a given system~\cite{Brunner2014}. We follow Refs.~\cite{Mermin1990,Zukowski2002} and write the general Bell inequality for an $N$ qubit configuration~\cite{Zukowski2002}:
\begin{eqnarray}\label{eq:zeta}
    \zeta &\leq &2^{-N}\sum_{s_1,...,s_N=\pm 1}\bigg|\sum_{k_1,...,k_N=1,2}\prod_{j=1}^N s_j^{k_j-1} E(\{k_j\})\bigg|,\nonumber\\
\end{eqnarray}
with $E(\{k_j\})$ defined in Appendix~\ref{app:app3}. Any value $\zeta>1$ that violates the inequality signals the presence of purely quantum correlations unable to be explained by local realism. For any set of recorded measurements, general Bell inequalities can be obtained employing linear programming techniques~\cite{Szangolies2017}.

We consider the set of operators $A=\{A_0,A_1\}$ and $B=\{B_0,B_1\}$ with:
\begin{eqnarray}\label{eq:bell_ops}
A_0 &=&I\otimes \sigma^z,\hspace{5pt} A_1=I\otimes \sigma^x,\nonumber\\
B_0 &=&I\otimes \left(\frac{-\sigma^z -\sigma^x}{\sqrt{2}}\right),\hspace{5pt} B_1=I\otimes \left(\frac{\sigma^x-\sigma^z }{\sqrt{2}}\right),
\end{eqnarray}
where $I$ represents the identity operator in the spatial subspace $\mathcal{S}$. Such a test belongs to the class of Bell inequalities for higher-dimensional systems~\cite{Cabello2002,Collins2002}. To perform the Bell test, we assign the set $A$ to $m$ particles and the set $B$ to $N-m$ particles. For $N=4$, we set $m=2$. We refer to Appendix~\ref{app:app3} for additional details on these expressions.

\section{Results}\label{sec:results}

\begin{figure}[ht!]
    \centering
    \includegraphics[scale=0.72]{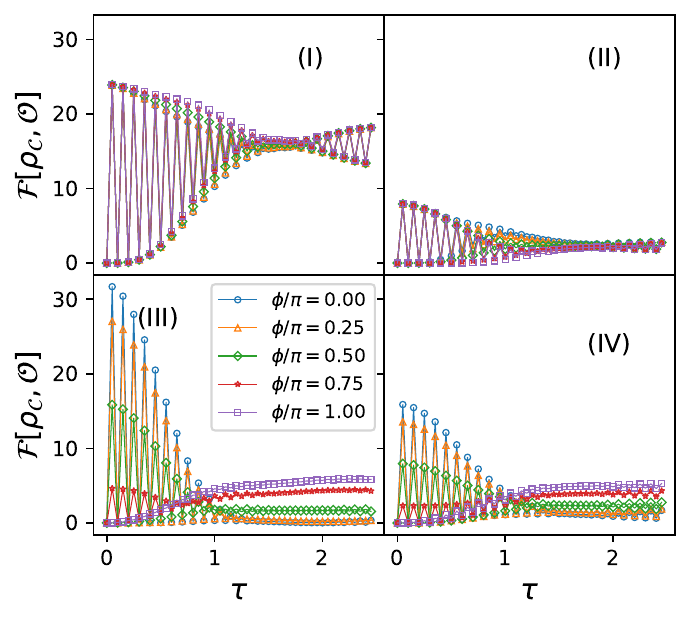}
    \caption{Quantum Fisher Information from Eq.~\eqref{eq:qfi_def} for the initial conditions in Fig.~\ref{fig:fig1}, for the state $\rho_{\mathcal{C}}$ defined in Eq.~\eqref{eq:rhoc_state}. For (II), particles with $\phi=0$ retain a higher value for the QFI than particles with $\phi=\pi$. In contrast, in configurations (III) and (IV) evolution of the QW leads to an increment in the QFI for particles with $\phi=\pi$, whereas for particles with $\phi=0$ the QFI tends to decrease. Data corresponds to $t=0.05$, a total of $50$ time steps, and $L=40$.  }
    \label{fig:fig4}
\end{figure}

\begin{figure}[ht!]
    \centering
    \includegraphics[scale=0.69]{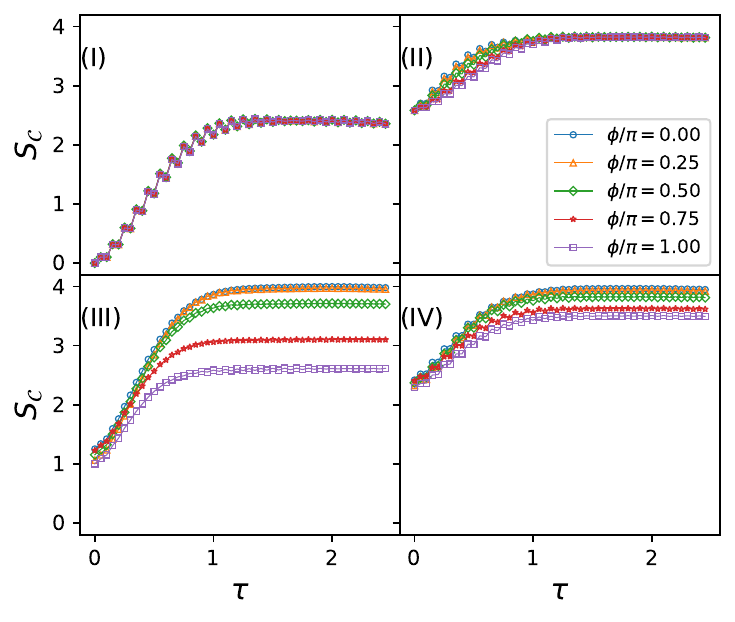}
    \caption{Entanglement entropy from Eq.~\eqref{eq:vn_entropy} for the initial conditions in Fig.~\ref{fig:fig1} for the state $\rho_{\mathcal{C}}$ defined in Eq.~\eqref{eq:rhoc_state}. Note that states (II), (III) and (IV) have initially significant entanglement. The von Neumann entanglement entropy grows with increasing the number of steps, signaling thermalizing behavior of the internal degrees of freedom with the environment. Note that specially for cases (III) and (IV), statistics of particles play a significant role in the reach value. Data corresponds to $t=0.05$, a total of $50$ time steps, and $L=40$.}
    \label{fig:fig5}
\end{figure}

In Fig.~\ref{fig:fig2}, we have represented the probability for two-particle coincidence events Eq.~\eqref{eq:diag_prob}, for the four different initial configurations in Fig.~\ref{fig:fig1} and different values of $\phi$. As expected, particles with $\phi=0$ have a higher probability of coincidence than particles with $\phi=\pi$, since the later experience Pauli exclusion principle. Particles with $\phi=0$ experience pronounced peaks due to boson bunching. We observe that for configuration (IV), the coincidence probability for a pair of fermions with $\phi=\pi$ always decreases for this configuration, contrary to (I) and (II) where peaks develop. For configuration (III), all $\phi$ values show a similar decay of the coincidence rates, with $\phi=\pi$ showing a slightly faster decay.

In Fig.~\ref{fig:fig3}, we have extracted through linear fitting the slope $v_\phi$ defined in Eq.~\eqref{eq:vf_phi}, for different values of $\phi$ and the initial configurations of Fig.~\ref{fig:fig1}~\footnote{In all fittings, we have checked that the average fitting error does not exceed $10^{-7}$. The fitting is performed in the middle region of the full time evolution of $\Delta$, taking the values of $\Delta_{12}$ in the interval $\tau\in\left[4.8,5.15\right]$ to avoid seeing boundary effects, staying within the linear regime (see Fig.~\ref{fig:app2})}. We observe that the statistics of particles influence the repulsion speed considerably in all configurations, with fermions $\phi=\pi$ presenting the largest separation speeds in all cases. Interestingly, configurations (I) and (IV) [(II) and (III)] present a similar dependence with the $\phi$ angle. We stress that $v_\phi$ is a joint property of the particles.

\begin{figure}[!t]
    \centering
    \includegraphics[scale=0.69]{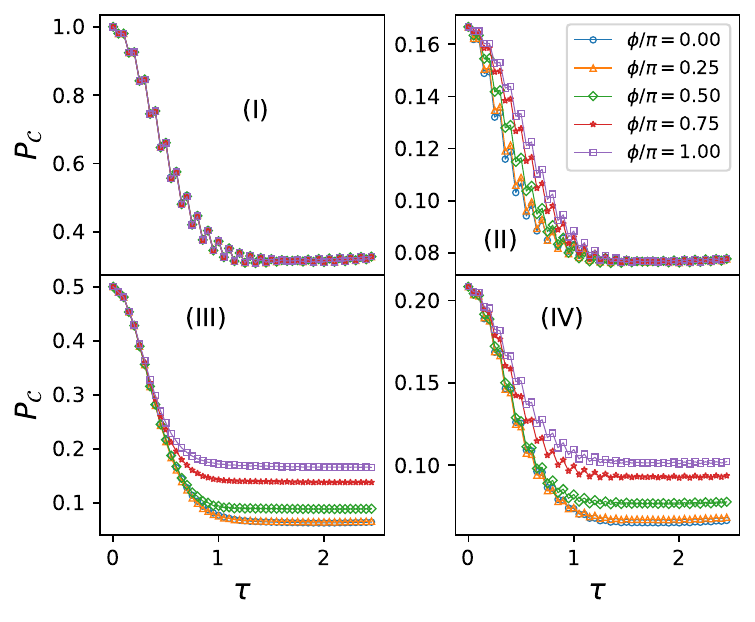}
    \caption{Purity from Eq.~\eqref{eq:purity}, for the initial conditions in Fig.~\ref{fig:fig1}, for the state $\rho_{\mathcal{C}}$ defined in Eq.~\eqref{eq:rhoc_state}. In accordance with Fig.~\ref{fig:fig5}, thermalizing behavior for bosons with $\phi=0$ leads to a maximally mixed (classical) state, whereas for fermions $\phi=\pi$ the purity remains above the maximally mixed state bound $P_\mathcal{C}=1/16$. Note the different vertical axes values. Data corresponds to $t=0.05$, a total of $50$ time steps, and $L=40$.  }
    \label{fig:fig6}
\end{figure}

In Fig.~\ref{fig:fig4} we represent for each of the initial configurations the QFI defined by Eq.~\eqref{eq:qfi_def}, corresponding to the state defined in Eq.~\eqref{eq:rhoc_state} and operator Eq.~\eqref{eq:hop_qfi}. We observe that while configurations (I) and (II) present quite similar results for different values of $\phi$, configurations (III) and (IV) show that while for $\phi=0$ the QFI oscillates evolving towards a zero value, for $\phi=\pi$ we find monotonic increment towards finite values. This indicates that for the Hamiltonian in Eq.~\eqref{eq:hop_qfi}, fermions are better suited to perform quantum metrology tasks after several time steps during the walk, given that the QFI sets a lower bound on the achievable precision for parameter estimation employing quantum states~\cite{Braunstein1994}. The results show that in the internal degrees of freedom subspace, multipartite entanglement increases (decreases) for fermions (bosons) during the evolution of the walk. 

In Fig.~\ref{fig:fig5} we have represented the von Neumann entanglement entropy as defined in Eq.~\eqref{eq:vn_entropy}. We observe that in all cases the entanglement between spatial and spin degrees of freedom characterized by $S_{\mathcal{C}}$ increases monotonically after several steps of the QW, eventually saturating its value. The effect of particle statistics is more pronounced in configurations (III) and (IV), where particles having fermion statistics ($\phi=\pi$) tend to become less entangled with their spatial degrees of freedom than particles possessing boson statistics ($\phi=0$), signaling absence of thermalization for the fermions in these cases. We stress that thermalization in this context refers to the amount of entanglement between the internal and the spatial degrees of freedom (i.e. the environment). Bosons tend to occupy equal spatial modes, eventually reaching a completely random reduced density matrix, i.e. a maximally mixed state, which is a classical state. On the other hand, 
fermions are restricted to the Pauli exclusion principle, which effectively reduces the dimensionality of 
available trajectories. It is therefore expected that their reduced density matrix is not a maximally mixed state.

The purity for the different initial configurations is represented in Fig.~\ref{fig:fig6}. In accordance with results of Fig.~\ref{fig:fig5}, fermions reach less mixing with their environment than bosons. This can be intuitively understood from the Pauli exclusion principle, which precludes equal spin occupation on a single spatial mode for fermions, thus effectively reducing the environment in which their internal degrees of freedom are embedded. This effect is expected to become more visible if more than one particle occupies the same spatial mode, as it is the case in (III) and (IV), where results suggest a complete absence of thermalization for $\phi\geq \pi/2$. This behavior is in stark contrast with what is observed in Fig.~\ref{fig:fig3}, which shows that the average interparticle distance speed is always higher for particles having closer to fermion statistics than for particles closer to bosons. Intuitively, a higher interparticle separation speed is associated with a faster thermalization rate over spatial modes; yet, results in Fig.~\ref{fig:fig4} suggest the opposite. This absence (presence) of thermalization for $\phi=\pi (\phi=0)$ is also manifesting in the amount of multipartite entanglement developing (decaying) through the QFI in Fig.~\eqref{fig:fig4}, since fully thermalized states converge to a classical mixture of states.

\begin{figure}[!t]
    \centering
    \includegraphics[scale=0.73]{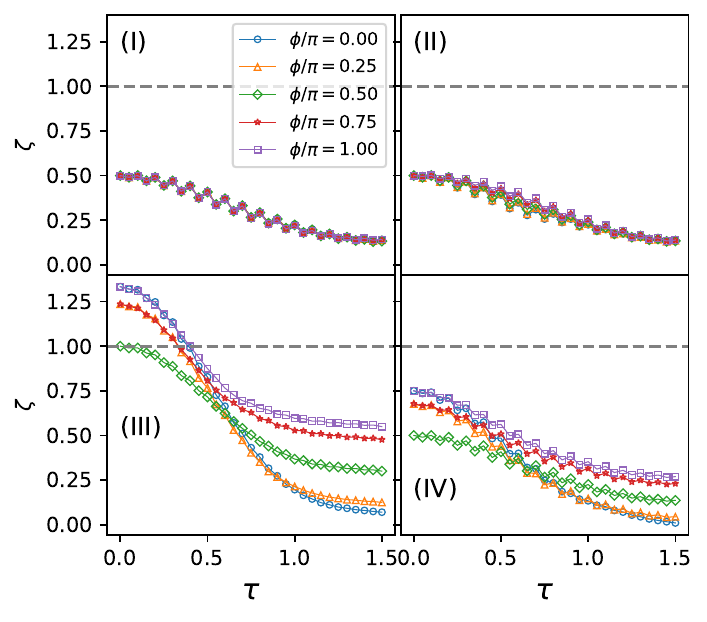}
    \caption{Performed Bell tests for the four different initial conditions in Fig.~\ref{fig:fig1}. Only condition (III) shows a violation of the Bell inequality in Eq.~\eqref{eq:bell_ineq} during the initial time steps, except for $\phi=\frac{\pi}{2}$, where the inequality is satisfied. The classical bound at $\zeta=1$ has been represented by a dashed grey line. Data correspond to $30$ time steps, $t=0.05$ and $L=40$.  }
    \label{fig:fig7}
\end{figure} 


Finally, we present results on the performed Bell tests in Fig.~\ref{fig:fig7}. We observe that configurations (I), (II) and (IV) have $\zeta<1$ at any times, and therefore can be in principle described by local realism. Configuration (III) violates the general Bell inequality in Eq.~\eqref{eq:bell_ineq} during the initial steps of the walk (except for $\phi=\pi/2$). Configuration (I) corresponds to a product state between internal degrees of freedom and the spatial degrees of freedom, rendering the state in the internal degrees of freedom purely classical. Configuration (II) can be regarded as a particular instance of two (I) configurations with $N=2$ each having opposite spins, therefore a local realism description is also expected in this case. Configuration (III) constitutes a pair of Bell states distantly separated from each other, constituting an initial quantum state with no classical description of correlations, except for $\phi=\frac{\pi}{2}$ where $\zeta=1$ initially. Finally, configuration (IV) consists of a Bell state and two classical separated spins. In that case, performing a Bell test would yield a combination from a maximal violation of the inequality and the most classical value of $\zeta$, resulting on average in a value below the $\zeta=1$ classical bound.

\section{Summary and outlook}\label{sec:conclusion}

In this work, we have investigated a multiparticle, discrete time QW on a 2D square lattice geometry, employing a system of $N$ identical particles prepared on different initial configurations. By tuning a single parameter $\phi$ associated with the particle statistics, we have reported results in joint properties and entanglement witnesses that could be potentially addressed in linear optics experiments employing multipartite entanglement between distinguishable photons as described in Sec.~\ref{subsec:multi_gene}. Our work lies in the context of multiparticle interference properties and the possibility to employ QWs of correlated photons~\cite{Matthews2013,Carolan2014,Jayakumar2014,Menssen2017,Lubasch2018} to simulate these processes. The results reported here could be explored on current integrated photonic circuits, complementing theoretical grounds for photonic quantum simulators.

Through the QFI we have shown that certain initial configurations of the particles allow to prepare states that are relevant for quantum metrology when the statistical nature of the particles is taken into consideration. In particular, for initial states sharing spatial modes between the particles, we report fermions to show a lower precision bound than bosons for parameter estimation of the all-to-all Ising model Hamiltonian. These results share context with recent works on the QFI in 1D continuous QWs of identical particles~\cite{Cai2021}. In the context of system-environment thermalization, we have shown that the particle statistics have a strong effect under certain initial preparations, concluding that contrary to bosons with $\phi=0$, fermions with $\phi=\pi$ do not always thermalize with their spatial degrees of freedom. Additionally, we have also investigated the nature of quantum correlations in the internal degrees of freedom, and whether these violate the general Bell inequality for $N$ qubits depending on the exchange statistics of particles. 

There are several questions left open for future research. These include employing different model realizations for the unitary evolution, alternative coin operators, different lattice geometries and initial configurations, or introducing interactions between particles. There is an evident connection of the model presented in this work with split-step models known to be relevant for exploring topological phases employing QWs~\cite{Kitagawa2010,Edge2015}. Identifying the associated split-step model for the QW presented here while investigating the presence or absence of nontrivial topological phases also constitutes an interesting avenue for future research. From the quantum metrology perspective, it is of interest to determine the role played by particle statistics during the evolution of the QW when more general operators $\mathcal{O}$ are employed in Eq.~\eqref{eq:qfi_def}. 

Our results address fundamental properties about the nature of quantum correlations in ensembles of identical particles subjected to a QW, particularly regarding entanglement properties in the context of system-bath thermalization of the internal degrees of freedom into spatial degrees of freedom. Employing photonic quantum simulators to engineer states that exploit particle statistics and entanglement to resist thermalization constitutes an interesting avenue for future research.

The data supporting the findings of this study are available from Zenodo~\cite{camacho_2025_15260434} and also on request from the corresponding author.

\section*{Acknowledgments}
We acknowledge discussions with Peter K. Schuhmacher, Jochen Szangolies and Helen M. Chrzanowski. This project was partially funded by the DLR ELEVATE project. We acknowledge funding by the Federal Ministry of Education and Research (BMBF) under grant number 13N15870.
\newpage
\bibliography{bib}

\appendix

\section{Eigenstates of the unitary operator}
\label{app:app1}
The Hamiltonians in Eq.~\eqref{eq:hams} can be both regarded as sets of parallel tight binding chains along the vertical and horizontal directions of the lattice. Thus, they have a diagonal representation in momentum $\vec{k}=(k_x,k_y)$ space:
\begin{eqnarray}
H_\parallel &=&\sum_{k_x,k_y}\varepsilon_{k_x}|k_x,k_y\rangle\langle k_x,k_y|,\nonumber\\
H_\perp &=&\sum_{k_x,k_y}\varepsilon_{k_y}|k_x,k_y\rangle\langle k_x,k_y|,
\end{eqnarray}
where we used that a lattice coordinates state with $\vec{r}=(x,y)$ can be expanded as:
\begin{eqnarray}
|x,y\rangle=\frac{1}{L}\sum_{\vec{k}}e^{i\vec{k}.\vec{r}}|\vec{k}\rangle,
\end{eqnarray}
and where we defined:
\begin{eqnarray}
\varepsilon_{k_{x,y}}=+2t\cos(k_{x,y}), \hspace{5pt}k_{x,y}&=&\frac{2\pi n_{x,y}}{L}\nonumber\\
n_{x,y}&=&0,1,...,L-1,
\end{eqnarray}
assuming periodic boundary conditions along the $x,y$ directions. The unitary operator consists of two terms:
\begin{eqnarray}
U&=&e^{-\ii H_{\parallel}}\otimes\left(\begin{matrix}
\cos\theta && \sin\theta\\
0 && 0
\end{matrix}\right) \nonumber\\
&+& e^{-\ii H_{\perp}}\otimes\left(\begin{matrix}
0 && 0\\
\sin\theta && -\cos\theta
\end{matrix}\right).
\end{eqnarray}
Note that $U$ is block diagonal in the $|\vec{k},\sigma\rangle$ basis, therefore each fixed $\vec{k}$ subspace can be diagonalized independently. Since $H_\parallel,H_\perp$ are diagonal in $\vec{k}$ space, we write for a fixed $\vec{k}=(k_x,k_y)$ the state:
\begin{eqnarray}
|\lambda_{\vec{k}}\rangle = \sum_{\alpha=\uparrow,\downarrow}\Phi(\vec{k},\sigma)|\vec{k},\sigma\rangle. 
\end{eqnarray}
Applying the unitary to $|\lambda_{\vec{k}}\rangle$:
\begin{eqnarray}
U|\lambda_{\vec{k}}\rangle&=&\Phi(\vec{k},\uparrow)\ e^{-\ii H_\parallel}\cos\theta|\vec{k},\uparrow\rangle  \nonumber\\
&+&\Phi(\vec{k},\downarrow)\ e^{-\ii H_\parallel}\sin\theta|\vec{k},\uparrow\rangle\nonumber\\
&-&\Phi(\vec{k},\downarrow)\ e^{-\ii H_\perp}\cos\theta|\vec{k},\downarrow\rangle  \nonumber\\
&+&\Phi(\vec{k},\uparrow)\ e^{-\ii H_\perp}\sin\theta|\vec{k},\downarrow\rangle 
\end{eqnarray}
We look now for solutions of the equation:
\begin{eqnarray}
U_{\vec{k}}\left(\begin{matrix}
\Phi(\vec{k},\uparrow)\\
\Phi(\vec{k},\dss)
\end{matrix}\right)&=&\lambda_{\vec{k}}\left(\begin{matrix}
\Phi(\vec{k},\uss)\\
\Phi(\vec{k},\dss)
\end{matrix}\right)
\nonumber\\
U_{\vec{k}}&=&\left(\begin{matrix}
e^{-\ii\varepsilon_{k_x}}\cos\theta && e^{-\ii\varepsilon_{k_x}}\sin\theta\\
e^{-\ii\varepsilon_{k_y}}\sin\theta && -e^{-\ii\varepsilon_{k_y}}\cos\theta
\end{matrix}\right).
\end{eqnarray}
The resulting eigenvalue equation is:
\begin{eqnarray}
\lambda_{\vec{k}}^2-\lambda_{\vec{k}}\cos\theta (e^{-\ii\varepsilon_{k_x}}-e^{-\ii\varepsilon_{k_y}})-e^{-\ii(\varepsilon_{k_x}+\varepsilon_{k_y})}=0.\nonumber\\
\end{eqnarray}
The eigenvalues are:
\begin{eqnarray}
\lambda_{\vec{k}}^\pm &=& \tau(\theta,\vec{k})\pm\sqrt{e^{-\ii(\varepsilon_{k_x}+\varepsilon_{k_y})}+\tau(\theta,\vec{k})^2},\nonumber\\
\tau(\theta,\vec{k})&=&\frac{\cos(\theta)\left(e^{-\ii\varepsilon_{k_x}}-e^{-\ii\varepsilon_{y}}\right)}{2}.
\end{eqnarray}
Since each $U_{\vec{k}}$ is unitary, these eigenvalues can be expressed as a phase $\lambda_{\vec{k}}^\pm=e^{\ii\varphi_{\vec{k}}^\pm}$. 
The eigenvector components satisfy:
\begin{eqnarray}
\Phi_\pm(\vec{k},\uss)=\frac{\sin\theta \Phi(\vec{k},\dss)}{ \lambda_{\vec{k}}^\pm e^{\ii\varepsilon_{k_x}}-\cos\theta}.
\end{eqnarray}
Normalizing we obtain the eigenstates for a fixed $\vec{k}=(k_x,k_y)$ value:
\begin{eqnarray}\label{eq:eigvecs_u}
|\lambda_{\vec{k}}^\pm\rangle &=&\left(\begin{matrix}
\frac{\sin\theta}{\sqrt{2-2\cos\theta\cos(\varphi_{\vec{k}}^\pm + \varepsilon_{k_x})}}\\
\frac{e^{\ii(\varphi_{\vec{k}}^\pm+\varepsilon_{k_x})}-\cos\theta}{\sqrt{2-2\cos\theta\cos(\varphi_{\vec{k}}^\pm + \varepsilon_{k_x})}}
\end{matrix}\right)=\left(\begin{matrix}
u_{\vec{k}}^\pm\\
v_{\vec{k}}^\pm
\end{matrix}\right).
\end{eqnarray}
Note that the above decomposition corresponds to a linear combination in the $\{|\vec{k},\uparrow\rangle,|\vec{k},\dss\rangle\}$ basis with coefficients $u_{\vec{k}}^\pm,v_{\vec{k}}^\pm$:
\begin{eqnarray}
|\lambda_{\vec{k}}^\pm\rangle=u_{\vec{k}}^\pm |\vec{k},\uss\rangle + v_{\vec{k}}^\pm|\vec{k},\dss\rangle.
\end{eqnarray}
Due to having distinct eigenvalues and $\hat{U}_{\vec{k}}$ being unitary, these eigenvectors are orthogonal to each other. 

If we evolve an initial state $|\psi_0\rangle=|\vec{r}_0,\sigma\rangle$ by the unitary $U^n$, we obtain:
\begin{eqnarray}
    U^n|\psi_0\rangle&=&\frac{1}{L}\sum_{\vec{k},\tau=\pm}e^{\ii n\lambda_{\vec{k}}^\tau + \ii\vec{k}.\vec{r}_0}\left(u_{\vec{k}}^{\tau}\delta_{\sigma,\uparrow} +v_{\vec{k}}^{\tau}\delta_{\sigma,\downarrow}\right)|\lambda_{\vec{k}}^\tau\rangle.\nonumber\\
\end{eqnarray}
\subsection{Two-particle return amplitude}
Having full knowledge of the eigenvectors and eigenvalues of the walk, we can obtain amplitudes from arbitrary initial state configurations. As an example, consider the two-particle QW given by the initial state:
\begin{eqnarray}\label{eq:psi0_bell}
    |\Psi_0\rangle=\frac{1}{\sqrt{2}}\left(|\vec{r}_0,\uparrow\rangle\otimes|\vec{r}_0,\downarrow\rangle + e^{\ii\phi}|\vec{r}_0,\downarrow\rangle\otimes |\vec{r}_0,\uparrow\rangle\right).\nonumber\\
\end{eqnarray}
We calculate the return amplitude:
\begin{eqnarray}
    A_0=\langle\Psi_0|\hat{U}^n|\Psi_0\rangle.
\end{eqnarray}
This amplitude can be expressed in terms of the coefficients:
\begin{eqnarray}
    A_0&=& \frac{1}{L^4}\left(\sum_{\vec{k},\tau}|u_{\vec{k}}^\tau|^2 e^{\ii n\lambda_{\vec{k}}^\tau}\right)\left(\sum_{\vec{k},\tau}|v_{\vec{k}}^\tau|^2 e^{\ii n\lambda_{\vec{k}}^\tau}\right)\nonumber\\
    &+&\frac{\cos(\phi)}{L^4}\left(\sum_{\vec{k},\tau}u_{\vec{k}}^\tau v_{\vec{k}}^{*,\tau} e^{\ii n\lambda_{\vec{k}}^\tau}\right)\left(\sum_{\vec{k},\tau}u_{\vec{k}}^{*,\tau} v_{\vec{k}}^{\tau} e^{\ii n\lambda_{\vec{k}}^\tau}\right).\nonumber\\
\end{eqnarray}
The second term is the only one depending on the particle statistics parameter $\phi$. Since $u_{\vec{k}}^\tau$ and $v_{\vec{k}}^\tau$ correspond respectively to spin up and down coefficients in the eigenstates Eq.~\eqref{eq:eigvecs_u}, particle statistics only affect transition amplitudes between these sectors for the state in Eq.~\eqref{eq:psi0_bell}.  

\section{Two-particle correlations}

\begin{figure}[!t]
    \centering
    \includegraphics[scale=0.58]{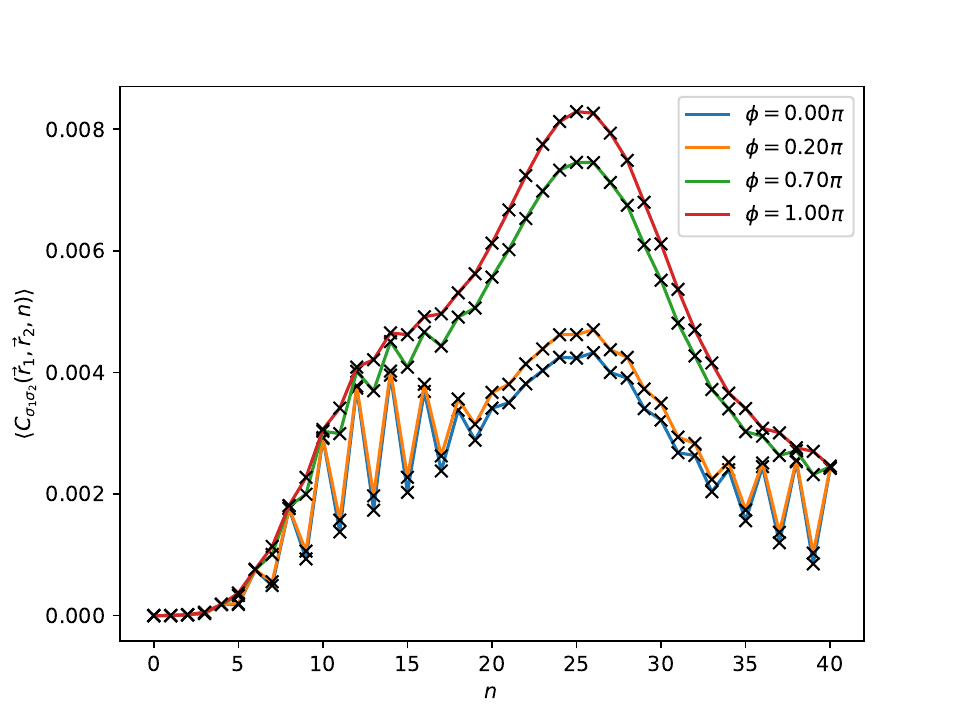}
    \caption{Benchmarked calculation for two-particle correlations on a square lattice of $L\times L$ dimensions, with $L=3$, $t=0.1$, $N=4$ particles and 40 total steps, for different values of $\phi$ in the initial state. The solid lines correspond to the exact numerical calculation by acting directly with the operators on the full state of dimension $(2L^2)^N$, whereas crosses indicate the values calculated with Eq.~\eqref{eq:C12_calc} using the single-particle wavefunctions coefficients. The $N=4$ particles are initialized at $(x,y,\sigma)\equiv(0,0,\uss),(0,2,\dss),(2,0,\dss),(2,2,\uss)$. The correlation is calculated for $\sigma_1,\sigma_2=\uss,\dss$ and $\vec{r}_1=\vec{r}_2=(2,2)$.
    }
    \label{fig:app1}
\end{figure}

\begin{figure}[!t]
    \centering
    \includegraphics[scale=0.64]{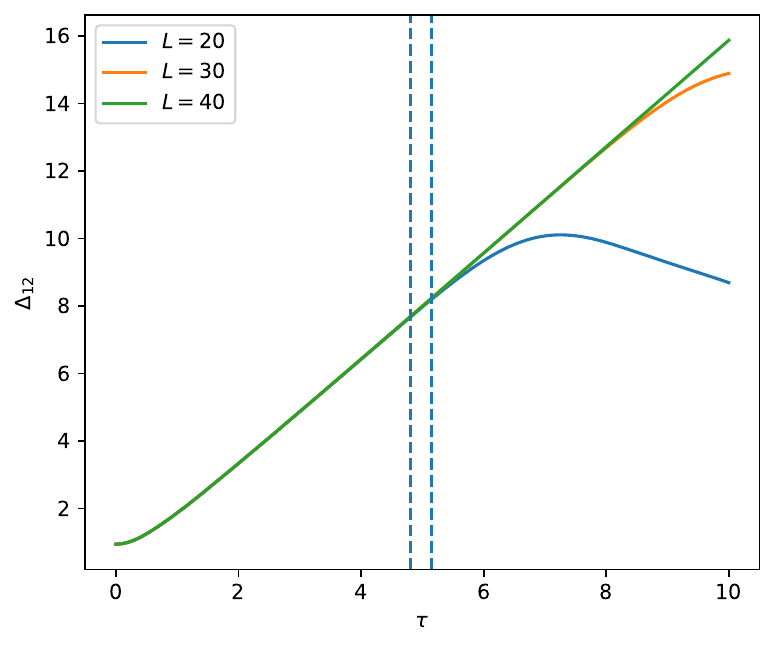}
    \caption{System size scaling of the inter-particle distance $\Delta_{12}$ for the quantum walk with initial configuration given by (III) in Fig.~\ref{fig:fig1}, with $\phi=\pi$, $\tau=nt$ with $t=0.05$. The deviation from linear behavior observed for smaller system sizes corresponds to boundary effects from particles reaching the edges of the lattice. The slope of $\Delta_{12}$ in the linear behavior region determines the propagation speed $v_\phi$. The slope $v_\phi$ is extracted in the region delimited by the dashed lines.
    }
    \label{fig:app2}
\end{figure}

\label{app:app2}
Since all particles forming the state $|\Psi(n)\rangle$ are identical, we can select any pair of subspaces to carry out the computation of the two-particle correlations (coincidence events). For instance, consider single-particle projectors $\hat{P}_1,\hat{P}_2$ with $\hat{P}_1\equiv \hat{P}_{1,\vec{r}_1,\sigma_1},\hat{P}_2 \equiv \hat{P}_{2,\vec{r}_2,\sigma_2}$. These operators act on the subspaces $\mathcal{H}_1,\mathcal{H}_2$, and thus correspond to operators that are local from the distinguishable particles point of view. The explicit formula for correlations after unitary evolution $U$ reads:
\begin{eqnarray}\label{eq:c12app}
C_{12}(n) &=& \sum_{p,p'}A_{p,p'}\langle\psi_{1p'}|(U^{\dagger})^n \hat{P}_1 U^n|\psi_{1p}\rangle\nonumber\\
&\times&\langle\psi_{2p'}|(U^{\dagger})^n \hat{P}_2 U^n|\psi_{2p}\rangle\nonumber\\
&\times&\prod_{q=3}^{N}\langle\psi_{qp'}|\psi_{qp}\rangle, \hspace{5pt}A_{p,p'}=\frac{e^{\ii(\phi_p-\phi_{p'})}}{N!}.
\end{eqnarray}
In Eq.~\eqref{eq:c12app}, the index $p$ labels all possible permutations resulting from the set of initial states $S_0$ defined in Sec.~\ref{subsec:initial_state}. There is a total of $(N!)^2$ terms in the sum. The product of brackets will cancel many terms when all initial conditions in $S_0$ are orthogonal to each other. The values of $A_{p,p'}=\frac{1}{N!}$ come from a symmetric permutation of both the ket and bra. From the rest of particles, there are $(N-2)!$ combinations. From the two chosen subspaces, we have $2C_2^N$ possible combinations, where $C_2^N$ is the binomial coefficient. From the other particles permutations, there is an equal number of phases with 0 or $\phi$ terms.

For convenience, we change the notation now and call the combined ket of initial states from particles 1 and 2 $|l,k\rangle$, i.e. both $l,k$ label one of the $N$ initial conditions in the set $S_0$, $l$ corresponding to particle 1 and $k$ corresponding to particle 2 (see main text for details). The only possible non-zero overlaps occur when the bra $\langle l',k'|$ contains the same values as in $l,k$, i.e. $l',k'\in\{l,k\}$. We also note that for any pair $(l,k)$, there is either a 1 or a phase $e^{i\phi}$. Thus, we can write the correlations for arbitrary single particle indices $1$ and $2$ as:
\begin{eqnarray} \label{eq:C12_calc}
C_{12}(n)&=&\sum_{l,k,l',k'}w_{lkl'k'}\langle l'|(U^\dagger)^n \hat{P}_1 U^n| l\rangle \langle k'|(U^\dagger)^n \hat{P}_2 U^n| k\rangle,\nonumber\\
w_{lkl'k'}&=&\begin{cases}
f_N,\hspace{5pt}l=l',k=k',\nonumber\\
f_N e^{-\ii\phi},\hspace{5pt}(l,k)\neq (l',k')\land l<k\nonumber,\\
f_N e^{+\ii\phi},\hspace{5pt}(l,k)\neq (l',k')\land l>k,
\end{cases}\nonumber\\
f_N &=& \frac{1}{N(N-1)}.
\end{eqnarray}
We note that this is independent of any geometry of the lattice, and that for the case of a single-particle Hilbert space with dimension $d=2L^2$ (with states labelled by the $l,k$ indices), there is a total of:
\begin{eqnarray}
    \left(\begin{matrix}
    d\\
    N
    \end{matrix}\right)=\frac{d!}{N!(d-N)!}
\end{eqnarray}
possible choices of the initial configuration with different $(l,k)$ indices pairs. A benchmark calculation of Eq.~\eqref{eq:C12_calc} is
shown in Fig.~\ref{fig:app1}. This allows to obtain in an efficient way
joint properties like $\Delta_{12}$ while addressing large system sizes,
as shown in Fig.~\ref{fig:app2}.

Note that the total number of particles (walkers) influence is two-fold: it appears in the pre-factor, but also in the sum over the set of single-particle indices $(l,k,l',k')$ specifying each of the initial conditions. We also note that for $n=0$ (initial condition), $C_{12}(0)$ is independent of the phase $\phi$, since only combinations with $l=l',k=k'$ can occur ($\hat{P}_1$ and $\hat{P}_2$ are projectors on single-particle states). This in turn implies that at initialization, simultaneous measurement of two-particle coincidences is independent of the particles statistics, which is easily verified. 

We can define the quantity:
\begin{eqnarray}
C^F_{12}(n)=\langle\Psi(n)|\hat{P}_1|\Psi(n)\rangle \langle\Psi(n)|\hat{P}_2|\Psi(n)\rangle.
\end{eqnarray}
Subtracting $C^F_{12}(n)$ from Eq.~\eqref{eq:C12_calc} modifies only the diagonal weights as:
\begin{eqnarray}
\bar{w}_{lkl'k'}&=&\begin{cases}
\frac{1}{N^3-N^2},\hspace{5pt}l=l',k=k',\nonumber\\
f_N e^{-\ii\phi},\hspace{5pt}(l,k)\neq (l',k')\land l<k\nonumber,\\
f_N e^{+\ii\phi},\hspace{5pt}(l,k)\neq (l',k')\land l>k.
\end{cases}
\end{eqnarray}

In the case of initial product states (i.e. a state representing distinguishable particles), the two-particle correlation becomes:
\begin{eqnarray}
\bar{C}_{12}(n)=\langle l|(U^\dagger)^n \hat{P}_1 U^n| l\rangle \langle k|(U^\dagger)^n \hat{P}_1 U^n| k\rangle.
\end{eqnarray}
Since particles are distinguishable in that case, the choice of $(l,k)$ pairs from the set $S_0$ is relevant for each subspace and leads to different results.

\section{Bell inequality for $N$ qubits}\label{app:app3}
If we trace over the spatial degrees of freedom for each particle subspace, the system becomes equivalent to a system of $N$ spin-$1/2$ state, where particle statistics are contained in the resulting complex coefficients. The reduced density matrix at time step $n$ is:
\begin{eqnarray}\label{eq:redrho}
    \rho_n&=&\text{Tr}_{\vec{r}}\left(|\Psi(n)\rangle\langle\Psi(n)\right)|=\sum_{p,p'}\frac{e^{\ii(\phi_p-\phi_{p'})}}{N!}A_{p,p',n},\nonumber\\
    A_{p,p',n}&=&\rho_1^{pp'n}\otimes...\otimes \rho_N^{pp'n},\nonumber\\
    \rho_{j}^{pp'n}&=&\left(\sum_{\vec{r}}\phi^p_{n,\vec{r},\sigma_j}\phi^{*,p'}_{n,\vec{r},\sigma_j'}\right)|\sigma_j\rangle\langle\sigma_j'|.
\end{eqnarray}
Note that the $A_{p,p',n}$ carry projectors in the spin degrees of freedom for each subspace $j$, therefore they are operators of dimension $2^N\times 2^N$. Note the state in Eq.~\eqref{eq:redrho} will be in general a mixed state, because the spatial and spin subspaces $\mathcal{S},\mathcal{C}$ are mixed due to the initial entangled states and subsequent unitary evolution.

We consider the general expression for a Bell inequality of $N$ spin-$1/2$ systems following Ref.~\cite{Zukowski2002}. In particular, we consider a set of $A_j(n_{k_j})$ with $k_j=1,2$ for each particle subspace $j$. The averaged outcome of experiments is given by:
\begin{eqnarray}
    E(k_1,...,k_N)\coloneqq E(\{k_j\})=\langle \prod_{j=1}^N A_j(n_{k_j})\rangle.
\end{eqnarray}
The expectation value is taken respect to the \emph{reduced density matrix} in the spin subspaces of the model, i.e. it lives on the Hilbert space $\mathcal{C}^{\otimes N}$.

The general expression for Bell inequalities of $N$ qubits reads~\cite{Zukowski2002}:
\begin{eqnarray}\label{eq:bell_ineq}
    \sum_{s_1,...,s_N=\pm 1}\bigg|\sum_{k_1,...,k_N=1,2}\prod_{j=1}^N s_j^{k_j-1} E(\{k_j\})\bigg|\leq 2^N.
\end{eqnarray}
We restrict to the special case where (compare Eq.~\eqref{eq:bell_ops}):
\begin{eqnarray}
    A_j(n_1)&=&\sigma^z,\hspace{5pt}A_j(n_2)=\sigma^x,\hspace{5pt}j\leq N/2,\nonumber\\
    A_j(n_1)&=&-\frac{1}{\sqrt{2}}(\sigma_x+\sigma_z),\hspace{5pt}j>N/2,\nonumber\\
    A_j(n_2)&=&\frac{1}{\sqrt{2}}(\sigma_x-\sigma_z), \hspace{5pt}j>N/2.
\end{eqnarray}
Defining:
\begin{eqnarray}\label{eq:zeta}
    \zeta=2^{-N}\sum_{s_1,...,s_N=\pm 1}\bigg|\sum_{k_1,...,k_N=1,2}\prod_{j=1}^N s_j^{k_j-1} E(\{k_j\})\bigg|,\nonumber\\
\end{eqnarray}
the condition for local realism description of correlations becomes:
\begin{eqnarray}
    \zeta\leq 1.
\end{eqnarray}

\section{Comparison with a split-step quantum walk}\label{app:app4}

\begin{figure}[ht!]
    \centering
    \includegraphics[scale=0.69]{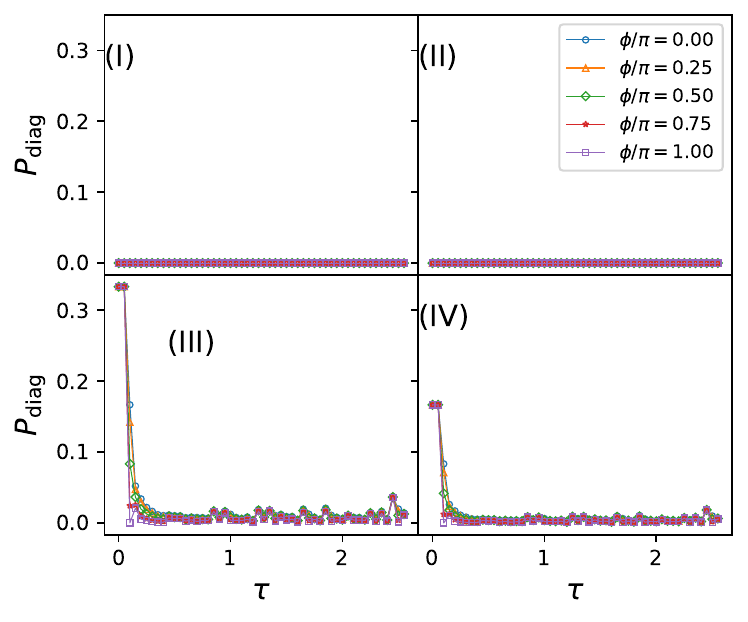}
    \caption{ Same figure as Fig.~\ref{fig:fig2} in the main text, for the model described in Eq.~\eqref{eq:u_st}. Data corresponds to a total of $50$ time steps, and $L=40$. Note that the time axes have been scaled by a factor $t=0.05$, with $\tau=nt$, for comparison purposes. }
    \label{fig:fig_app3}
\end{figure}

\begin{figure}[ht!]
    \centering
    \includegraphics[scale=0.69]{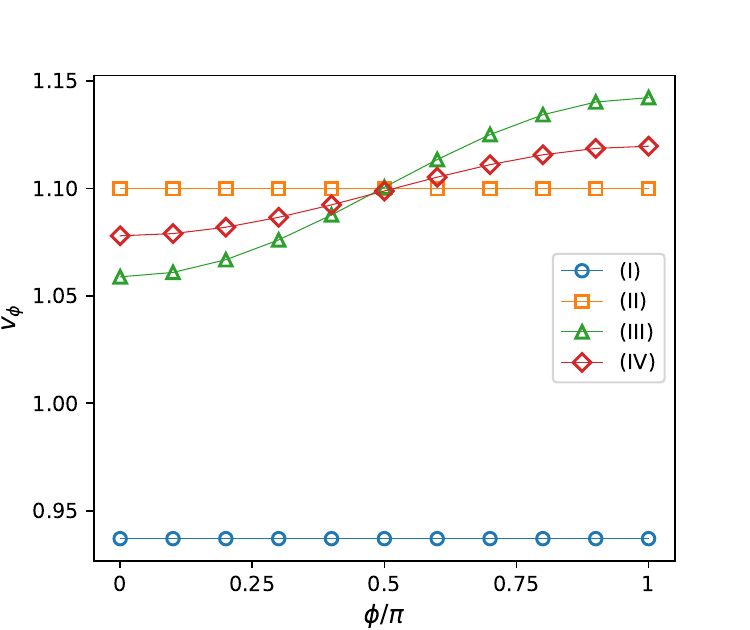}
    \caption{ Same figure as Fig.~\ref{fig:fig4} in the main text, for the model described in Eq.~\eqref{eq:u_st}. Data corresponds to a total of $12$ time steps, and $L=40$. In this case, no scaling of the $x$ axes was performed to extract the velocities, i.e. $v_\phi$ is obtained by setting $t=1$ in Eq.~\eqref{eq:vf_phi}. }
    \label{fig:fig_app4}
\end{figure}

In this Appendix, we show results when comparing the quantum walk model employed in this work with a split-step version of the quantum-walk models found in Refs.~\cite{Kitagawa2010,Edge2015}. We consider the unitary:
\begin{eqnarray}\label{eq:u_st}
\hat{\mathcal{U}} &=& \mathcal{U}^{\otimes N},\nonumber\\
\mathcal{U}&=&U_yU_{\mathcal{C}}U_{x}U_{\mathcal{C}},\nonumber\\
U_x &=&\sum_{x,y}\sum_{s=\pm 1} |x+s,y,\sigma_s\rangle\langle x,y,\sigma_s|,\nonumber\\
U_y &=&\sum_{x,y}\sum_{s=\pm 1} |x,y+s,\sigma_s\rangle\langle x,y,\sigma_s|,
\end{eqnarray}
where $U_\mathcal{C}$ is given in Eq.~\eqref{eq:uni_def} and we have defined $\sigma_{s=1}\equiv \uparrow$ and $\sigma_{s=-1}\equiv \downarrow$. We set $\theta=\pi/4$ and employ the initial configuration for the particles given in Fig.~\ref{fig:fig1}. Note that this model does not include a hopping parameter $t$, and that two coin operations are performed on a single application of the unitary $\hat{U}$ over the state (see Eq.(5) in Ref.~\cite{Edge2015} for a direct comparison). For simplicity on the implementation, we consider periodic-boundary conditions (PBC), which folds the spatial lattice into a torus geometry.

In Figs.~\ref{fig:fig_app3},~\ref{fig:fig_app4} we have represented the coincidence probabilities and the spread velocities, to allow for a direct comparison with figures~\ref{fig:fig2} and~\ref{fig:fig3} in the main text. The main thing to note is that for the model in Eq.~\eqref{eq:u_st}, particle statistics become irrelevant if particles do not share any spatial modes in the chosen initial configurations. On the contrary, configurations (III) and (IV) experience coincidence events at the initial steps, until particles start separating from each other. From Fig.~\ref{fig:fig_app4}, we conclude that for configurations (I) and (II), particles spread at equal paces independent on the statistics, in accordance with Fig.~\ref{fig:fig_app3}. Configurations (III) and (IV) have spread velocities $v_\phi$ dependent on the particle statistics parameter $\phi$, with fermions $(\phi=\pi)$ experiencing the fastest separation.  

Direct comparison with results obtained from the model in Eq.~\eqref{eq:uni_def} shows that introducing complex amplitudes (with small $t=0.05$) using the unitary evolution operator in Eq.~\eqref{eq:uni_def} leads to a richer dynamical structure of correlations. Indeed, the appearance of complex phases in the tunneling amplitudes leads to modifications in the relative phases present in the quantum state, resulting in emergent correlations not seen for the real-valued unitary operator in Eq.~\eqref{eq:u_st}.

\end{document}